\RequirePackage{fix-cm}
\documentclass[smallextended,natbib]{svjour3}       
\smartqed  

\usepackage{graphicx}

\usepackage{mathptmx}      

\usepackage{natbib}
\usepackage{amssymb,amsmath}
\usepackage{array}
\usepackage{multirow}
\usepackage{multicol}
\usepackage[]{booktabs}
\usepackage{tabularx}
\usepackage{complexity}
\usepackage{graphics}

\journalname{Optimization Letters}
\begin{document}

\title{A branch and cut algorithm for minimum spanning trees under conflict constraints
}


\author{Phillippe Samer \and Sebasti\'{a}n Urrutia}


\institute{Phillippe Samer \at
              Universidade Federal de Minas Gerais (UFMG), Belo Horizonte, MG, Brazil \\
              Tel.: +55-31-3409 5851\\
              Fax: +55-31-3409 5858\\
              \email{samer@dcc.ufmg.br}           
           \and
           Sebasti\'{a}n Urrutia \at
              Universidade Federal de Minas Gerais (UFMG), Belo Horizonte, MG, Brazil \\
              Tel.: +55-31-3409 7533\\
              Fax: +55-31-3409 5858\\
              \email{surrutia@dcc.ufmg.br} \\
}

\date{Received: 5 July 2013 / Accepted: 5 May 2014}

\maketitle

\begin{abstract}
We study approaches for the exact solution of the \NP--hard minimum spanning tree problem under conflict constraints. Given a graph $G(V,E)$ and a set $C \subset E \times E$ of conflicting edge pairs, the problem consists of finding a conflict-free minimum spanning tree, i.e. feasible solutions are allowed to include at most one of the edges from each pair in $C$. 
The problem was introduced recently in the literature, with several results on its complexity and approximability. Some formulations and both exact and heuristic algorithms were also discussed, but computational results indicate considerably large duality gaps and a lack of optimality certificates for benchmark instances.
In this paper, we build on the representation of conflict constraints using an auxiliary conflict graph $\hat{G}(E,C)$, where stable sets correspond to conflict-free subsets of $E$.
We introduce a general preprocessing method and a branch and cut algorithm using an IP formulation with exponentially sized classes of valid inequalities for both the spanning tree and the stable set polytopes.
Encouraging computational results indicate that the dual bounds of our approach are significantly stronger than those previously available, already in the initial LP relaxation, and we are able to provide new feasibility and optimality certificates.
\end{abstract}


\section{Introduction}
\label{sec:intro}

Disjunctive relations arise in many contexts of combinatorial optimization and integer programming (IP): different problems have been studied under conflict or multiple choice constraints, disjunctive cuts are implemented in several mixed-integer programming solvers,
and the Disjunctive Programming framework \citep{balasDP2010} pioneered by Egon Balas in the 1970s is still relevant and has relationships with other IP techniques.
Nevertheless, disjunctively constrained versions of classic problems in graph theory such as shortest paths, spanning trees and matchings were studied only recently \citep{damPTM2011}.

The literature on such problems on graphs regards mainly complexity and approximability results, but a particular interest in the minimum spanning tree problem under conflict constraints has led to the development of algorithms and benchmark instances \citep{Zhang2011}. Its feasibility version is also discussed in the context of the quadratic bottleneck spanning tree problem \citep{punnenQB2011}.

In this paper we discuss approaches for the exact solution of the minimum spanning tree problem under conflict constraints (MSTCC).
Given a graph $G(V,E)$ and a set $C \subset E \times E$ of conflicting edge pairs, the problem consists of finding a conflict-free minimum spanning tree (MST): a spanning tree of $G$, of minimum cost, which includes at most one of $e_i$ or $e_j$ for each pair $\left\lbrace e_i, e_j \right\rbrace \in C$.

An equivalent definition which we exploit here uses the concept of a conflict graph $\hat{G}(E,C)$: by denoting each edge in the original graph as a node in $\hat{G}$, we represent each conflict constraint by an edge connecting the corresponding nodes in $\hat{G}$. The problem is thus to find a subset of $E$ of minimum cost, corresponding both to a spanning tree of $G$ and to a stable set in $\hat{G}$.
While this auxiliary graph is introduced with the problem \citep{damPTM2011}, conflict graphs have been used for many years in integer programming to represent logical relations among variables \citep{AtamturkConflict2000} or, in a different sense, to leverage SAT conflict analysis techniques to generate cutting planes from pruned nodes in the enumeration tree \citep{satConflict2007}.

\paragraph{Related work}
\cite{mstccLNCS2009,damPTM2011} introduce problems on paths, trees and matchings under conflict and forcing constraints (the latter requires solutions to include at least one of the edges from each pair in $C$).
The authors establish several results on the complexity and approximation hardness of such problems.
They prove that MSTCC is strongly $\NP$--hard, even when every connected component of the conflict graph is a path of length two.
Moreover, it cannot be approximated by a constant factor of the optimal value, unless $\P=\NP$.

Both theoretical and computational results on MSTCC are further described by \cite{Zhang2011}. The authors discuss special cases which are polynomially solvable, feasibility tests, heuristics and two exact algorithms based on Lagrangean relaxation schemes. One formulation is integral, as all disjunctive constraints are relaxed and the classic MST is solved as subproblem. The other approach relaxes only part of the conflicts and solves the $\NP$--hard maximum edge clique partitioning as subproblem.
Computational results are discussed, but considerably large duality gaps are left by such algorithms, which do not provide optimality certificates.

Finally, we note that recent papers on different combinatorial problems under disjunctive constraints can be found.
\cite{pferschyKP2009} discuss the complexity of special cases of the knapsack problem with conflict constraints.
\cite{SadykovBPbranchAndPrice2012} describe a branch and price algorithm for the bin packing problem with conflicts, solving conflict-constrained knapsack subproblems with dynamic programming.
The complexity of maximum flow problems under conflict and forcing constraints has also been studied by \cite{flow2011journal}.
Polynomially solvable cases, a Lagrangean relaxation scheme and heuristics for minimum cost perfect matchings under conflict constraints are described by \cite{OncanMatch2013}.

\paragraph{Our contribution} 
We describe a preprocessing algorithm and a branch and cut approach for the solution of MSTCC, separating inequalities corresponding both to the spanning tree polytope (to guarantee the solution is an acyclic subgraph of the original graph $G$) and the stable set polytope (to tighten the representation of feasible node subsets in the conflict graph $\hat{G}$).
Encouraging results include stronger dual bounds already in the root LP relaxation, for all instances in a benchmark set.
Finally, we present several optimality certificates and new results for five instances whose feasibility was previously unknown: two of which are feasible, and three that are proved to be infeasible.


\section{A preprocessing algorithm}
\label{sec:preprocessing}

\begin{figure}[h]
\begin{center}
    \includegraphics[scale=0.61]{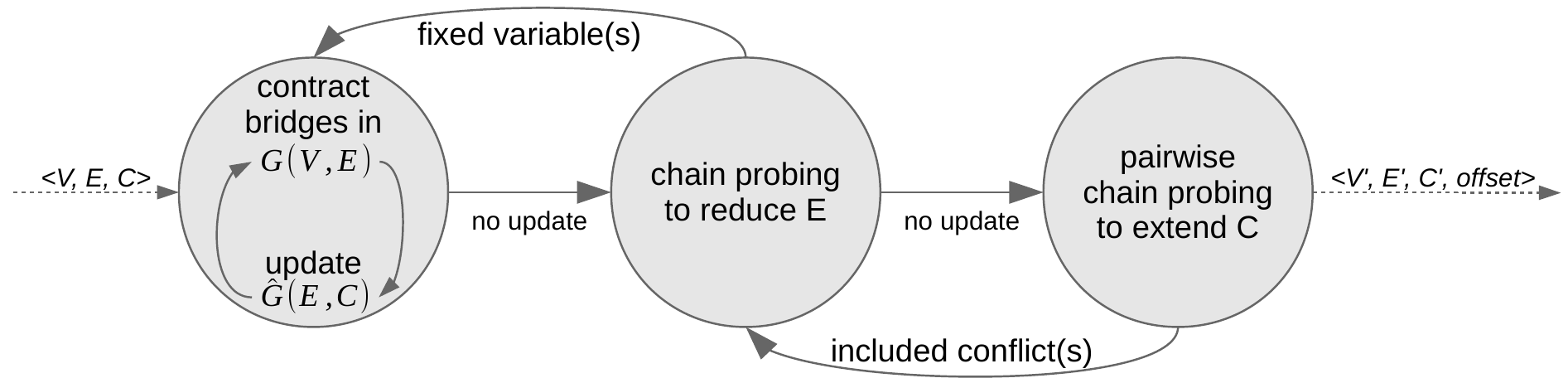}
\end{center}
\caption{Three steps of the preprocessing algorithm, given input graphs $G(V,E)$ and $\hat{G}(E,C)$. Both a contraction in the first step and a removal in the second implies fixing edge variables and updating the corresponding conflicting pairs in $C$. The output includes the objective function offset due to contracted edges.
}
\label{fig:preprocessing}
\end{figure}

As an enhancement to the solution process, we considered using MSTCC problem-specific feasibility conditions to devise a preprocessing algorithm. Figure \ref{fig:preprocessing} depicts the overall method, which we describe next. It is worth remarking that, although designed in the context of our branch and cut approach, the following algorithm can be integrated to any solution technique for the problem.

Let a MSTCC input instance consist of the original graph $G(V,E)$ and the conflict one $\hat{G}(E,C)$. The general algorithm is a three-phase iterative process, where each phase is executed as long as the problem instance is updated. To every step fixing an edge in $G$ corresponds an update in the conflicting pairs in $\hat{G}$.

The first phase checks for cut-edges (bridges) in $G$, using depth-first search. As long as the original graph is connected, any cut-edge $e_1$ is contracted, its cost $c_{e_1}$ is added as an offset to the optimal value of the reduced problem (if any), and conflicting pairs are removed both from $G$ and $\hat{G}$, i.e. we can fix variables corresponding to $e_k$ to zero, for all $e_k \in E$ such that  $\{e_1, e_k\} \in C$. If at any point we verify that $G$ is not connected, the original problem is infeasible;  on the other hand, if the resulting graph is a conflict-free tree, it is also the unique feasible solution to the problem.

The remaining phases use the \emph{probing} technique (i.e. evaluating the consequences of possibly setting a binary variable to one of its bounds), based on implications from feasibility conditions, as \cite{AtamturkConflict2000} denote. Nevertheless, in the context of that work it means analyzing the structure of a general IP to derive infeasibility implications, while our next steps analyze the combinatorial structure at hand: any solution is required to induce a connected subgraph of $G$, and conflicting edge pairs might render that infeasible after tentatively fixing variables in chain.

In the second phase, we check the connectivity of subgraphs of $G$ including a given edge $e$ (with degree in the conflict graph $\delta_{\hat{G}}(e) > 0$).
If the chain removing conflicting pairs and fixing any cut-edges possibly implied by the selection leads to a disconnected graph, we may remove $e$ from $E$ and the corresponding conflicts from $C$. In this case, we return to the first phase as $G$ might include new cut-edges.

Finally, if no edge could be fixed in the previous step, a third phase performs a similar evaluation on the connectivity of $G$, now probing pairs of variables. The chain starts fixing in the solution edges $e_1$ and $e_2$ (neither already in conflict with each other nor both conflict-free in $\hat{G}$), and proceeds by removing conflicting ones and including any cut-edges implied by the selection.
Now, if $G$ would become disconnected, the new conflict pair $(e_1, e_2)$ is included in $C$, and we may return to the second phase to check if it is possible to remove any edge.

If an iteration is completed without any update on the third phase, the reduced instance and the objective value offset are output. Note that, if any conflict-free spanning tree of $G$ is obtained during the last two phases, we may store it as a primal feasible solution.


\section{Branch and cut approach}
\label{sec:bnc}

We describe next the representations of the spanning tree and stable set polytopes on which we build the proposed formulation for MSTCC, and our branch and cut approach.
In the following, suppose we are given an input graph $G(V,E)$, with $|V| = n$, $|E| = m$, weights $c_e$ associated with each edge, and a set $C \subset E \times E$ of conflicting edge pairs.
Given a non-empty subset $S \subseteq V$, let $E(S) \subseteq E$ be the set of edges with both endpoints in $S$.
Define the incidence vector $\textbf{x} = (x_1, x_2, \dots, x_{|E|})$ of a given solution so that $x_e = 1$ if edge $e$ is included in the solution, and $x_e = 0$ otherwise.
Whenever possible, we use $P_{name}$ to denote $P_{name}(G)$, the polyhedron of interest considering a particular graph $G$.

\subsection{Integer programming formulation}
\label{sec:bnc:formulation}

Let $P_{sec} \subset \mathbb{R}_+^{|E|}$ denote the representation of the spanning tree polytope with subtour elimination constraints (SEC), given by:
\begin{alignat}{2}
& \sum_{e \in E(S)} x_e  \leq |S| - 1, 	\quad && \forall S \subset V, S \ne \emptyset  \label{mp:tree_sec1} \\
& \sum_{e \in E} x_e  = n-1                                                    \label{mp:tree_sec2} \\
& 0 \leq x_e \leq 1, 					\quad && \ e \in E                                                 \label{mp:tree_sec3}
\end{alignat}

\vspace{-0.1cm}
While SEC~(\ref{mp:tree_sec1}) enforce a cycle-free condition, since any connected subgraph on $S$ inducing a cycle has at least $|S|$ edges, a feasible solution is guaranteed to be a spanning tree of $G$ by picking $n-1$ edges~(\ref{mp:tree_sec2}). Constraints (\ref{mp:tree_sec3}) correspond to the continuous relaxation of binary variables $x_e$.

An important result by \cite{edmondsMatroidsGreedy1971} is that the above formulation is tight, as all its vertices are integer-valued.
Next, we want to describe the polyhedral region corresponding to conflict-free incidence vectors.
In this paper, we use a particular representation of the stable set polytope $P_{stab}$ to regard the feasibility of a solution with respect to the conflict graph $\hat{G}(E,C)$. 
Unfortunately, as the stable set problem is $\NP$-hard (and assuming $\P \ne \NP$), an ideal description of $P_{stab}$ is not available.

First, consider the stable set polytope relaxation $P_{rstab} \subset \mathbb{R}_+^{|E|}$, given by (\ref{mp:tree_sec3}) and:
\begin{alignat}{2}
& x_{e_1} + x_{e_2} \leq 1,\                    &\ & \hspace{1cm} \forall \{e_1, e_2\} \in C     \label{mp:stab1}
\end{alignat}
Edge inequalities (\ref{mp:stab1}) guarantee a conflict-free solution.
Together with non-ne\-ga\-ti\-vi\-ty inequalities in (\ref{mp:tree_sec3}), these are enough to formulate the simplest relaxation of $P_{stab}$, which yields the convex hull of stable sets in $\hat{G}$ if and only if the conflict graph is bipartite \citep{PadbergSurvey1979}.

We strengthen this representation with the intersection of two tighter polyhedra, as we describe next. The first is known as the cycle-constraint stable set polytope $P_{cstab} \subset P_{rstab} \subset \mathbb{R}_+^{|E|}$, given by (\ref{mp:tree_sec3}), (\ref{mp:stab1}) and odd-cycle inequalities:
\begin{alignat}{2}
& \sum_{i \in U} x_i  \leq \frac{|U| - 1}{2}, \ &\ & \hspace{0.1cm} \forall U \subset E \ \text{inducing an odd-cycle in } \hat{G}   \label{mp:stab2}
\end{alignat}
These are valid for $P_{stab}$ since the cardinality of any stable set in a subset $U$ of vertices of $\hat{G}$ inducing an odd-cycle (with or without chords) is at most $\lfloor \frac{|U|}{2} \rfloor = \frac{|U| - 1}{2}$.
Still, $P_{cstab}$ yields the convex hull of stable sets in $\hat{G}$ if and only if the conflict graph is \emph{t-perfect} \cite[Section 9.1]{GLSbook1988}. Although this is a quite restrictive condition, the separation of odd-cycle inequalities remarkably improves the quality of dual bounds for MSTCC benchmark instances, as we describe in Section \ref{sec:results}.

Finally, an additional relaxation consists of the clique-constraint stable set polytope $P_{qstab} \subset P_{rstab} \subset \mathbb{R}_+^{|E|}$, described by (\ref{mp:tree_sec3}) and clique inequalities:
\begin{alignat}{2}
& \sum_{i \in Q} x_i \leq 1,\                    &\ & \hspace{1cm} \forall Q \subset E \text{ inducing a clique in } \hat{G}    \label{mp:stab3}
\end{alignat}
The unit upper bound is clearly valid for $P_{stab}$, as no stable set in $\hat{G}$ could include more than one vertex from any complete subgraph.
The class of graphs for which $P_{stab}$ and $P_{qstab}$ coincide is precisely that of perfect graphs \citep[Section 9.2]{GLSbook1988}.

We remark that, when they induce odd-holes (chordless odd-cycles), inequalities (\ref{mp:stab2}) might define facets of $P_{stab}$, and the sequential lifting procedure of \cite{padberg1973} could be used for that purpose. Padberg also proved that inequalities (\ref{mp:stab3}) are facet-defining for $P_{stab}$ if and only if the clique induced by $Q$ is maximal. We build on this last result to include exactly those non-dominated inequalities in the model (see Section \ref{sec:bnc:algorithm}).
Unfortunately, while clique inequalities traditionally had a strong impact in solving stable set problems, the impact of these in MSTCC benchmark instances was less expressive, as we evaluate in Section \ref{sec:results}.

In conclusion, we formulate the MSTCC problem as
\begin{equation}
\min \left\lbrace \sum_{e \in E} c_e x_e : \textbf{x} \in P_{sec} \cap P_{stab} \subseteq \mathbb{B}^{|E|}  \right\rbrace
\label{mp:mstcc}
\end{equation}
and will therefore approximate $P_{stab}$ as $P_{cstab} \cap P_{qstab}$.

\subsection{General algorithm}
\label{sec:bnc:algorithm}

After preprocessing the problem instance with the algorithm described in Section \ref{sec:preprocessing}, we propose a branch and cut approach for MSTCC, solving (\ref{mp:mstcc}) with the generation of cutting planes corresponding to SEC (\ref{mp:tree_sec1}) and odd-cycle inequalities (\ref{mp:stab2}).

In general, the separation problem associated with clique inequalities (\ref{mp:stab3}) is \NP-hard; in fact, the optimization problem over $P_{qstab}$ itself is \NP-hard \citep[Section 9.2]{GLSbook1988}.
Nevertheless, we verified that conflict graphs in the set of challenging benchmark instances for MSTCC used in the literature have a limited number of maximal cliques: except for one instance, $\hat{G}(E,C)$ has less than $|C|$ maximal cliques, leading to a smaller model than using edge-inequalities (\ref{mp:stab1}). In fact, after the preprocessing phase, that was the case for the complete benchmark set.

We could therefore successfully use the maximal clique enumeration algorithm of \cite{tomita2006} to actually include the corresponding (non-dominated subset of) maximal clique inequalities \emph{a priori}.
It is extremely fast in practice, with negligible runtime for MSTCC instances. The authors also prove that the worst-case complexity of the algorithm ($O(3^{m/3})$, in an $m$-vertex graph) is optimal with respect to $m$, since there can be at most $T(m) \leq 3^{m/3}$ maximal cliques. We consider that our methodology could be used when $T(m)$ is in $O(|C|)$. 
Alternatively, when using the present formulation to solve a different instance set, which renders the enumeration method infeasible, one could use a greedy heuristic to: (i) replace edge-inequalities by any maximal clique containing it; (ii) lift any violated triangle identified during the separation procedure for odd-cycle inequalities to a larger clique (\cite{RebennackTutorial2012} suggests that it could be effective in the context of stable set instances).

Now, the algorithm starts with the solution of $\min \left\lbrace \sum_{e \in E} c_e x_e \right\rbrace$ subject to (\ref{mp:tree_sec2}), (\ref{mp:tree_sec3}) and (\ref{mp:stab3}).
Let \textbf{x} be the solution to such linear program (LP). Clearly, if \textbf{x} is integral and a depth-first search from any vertex $i \in V$ reaches every other vertex in $V\backslash \{i\}$, then \textbf{x} is also the optimal solution of the original IP (\ref{mp:mstcc}).

Otherwise, we search for SEC (\ref{mp:tree_sec1}) as well as odd-cycle inequalities (\ref{mp:stab2}) violated by \textbf{x}, which strengthen the relaxed polyhedron. This is performed by the separation procedures we describe next.
We check both classes of inequalities for violation at a given solution \textbf{x}.
If any procedure is able to separate \textbf{x}, we add the corresponding cuts globally, and solve the new, reinforced LP.
If both separation procedures fail to find a violated inequality, we branch on variables and iterate.

\subsection{Separation procedures}
\label{sec:bnc:separation}

We describe next the algorithms for solving the exact separation problem for SEC (\ref{mp:tree_sec1}) and for odd-cycle inequalities (\ref{mp:stab2}). Apart from details of our implementation, these are actually quite standard, and the reader is referred to classic expositions.

\subsubsection*{Subtour elimination constraints}

\cite{optimalTrees1995} describe a standard procedure to look for violated SEC in a solution \textbf{x} to the relaxation of $P_{sec}$.
First, a directed network corresponding to \textbf{x} is built, with the capacity of both arcs $(i,j)$ and $(j,i)$ set to the current value of edge variable $x_{i,j}$. We also set an arbitrary vertex as root $r$; we use $r=1$.
Now, \textbf{x} satisfies all subtour elimination constraints if and only if we can send one unit of flow from $r$ to every other vertex in the capacitated network.

Therefore, by performing $n-1$ maximum flow (minimum cut) computations, from $r$ to every vertex $i \in V \backslash \left\lbrace r \right\rbrace$, we may check in polynomial time if \textbf{x} is feasible in $P_{sec}$:
if the value of any minimum cut is less than $1$, we have found a violated inequality.
To find a minimum $(r,i)$ cut, we use an implementation of the highest-label preflow-push algorithm of \cite{GoldbergTarjan1988}, available in the open-source Library for Efficient Modeling and Optimization in Networks \citep{lemon2011}.

\subsubsection*{Odd-cycle inequalities}

\cite{GerardsSchrijver1986} introduce an exact separation procedure for odd-cycle inequalities, which is clearly described in the tutorial of \cite{RebennackTutorial2012}.
Much like the separation of SEC, we may check in polynomial time if every odd-cycle inequality is satisfied by computing a minimum cost cycle in an auxiliary graph.

We start by defining a new weight function $w$ for adjacencies in the conflict graph $\hat{G}(E,C)$: let $w(u,v) = \frac{(1 - x_u - x_v)}{2}$ for each $\{u,v\} \in C$. Since constraints (\ref{mp:tree_sec3}) and (\ref{mp:stab1}) are satisfied \emph{a priori}, we have that $w: C \rightarrow [0,\frac{1}{2}]$.
We also construct an auxiliary bipartite graph $H$ by duplicating $\hat{G}$ as follows: $H$ has two vertices $u^+$ and $u^-$ for each $u \in E$, as well as edges $\{u^+, v^-\}$ and $\{u^-, v^+\}$ for each $\{u,v\} \in C$, both with weight $w(u,v)$.

Then, for each $u \in E$, we compute a shortest $(u^+, u^-)$ path in the auxiliary graph $H$. Note that, as the weight function $w$ is non-negative, we may use Dijkstra's algorithm, stopping its execution as soon as the goal vertex $u^-$ is selected.
By the construction of $H$, the vertices $u^+$ and $u^-$ are in different sets of the bipartition, implying that the path has odd length. 
By omitting the $^+$ and $^-$ indices, the path corresponds to a closed odd-walk in $\hat{G}$. However, this walk might include repeated nodes and edges, since the shortest path is determined in $H$; in fact, $H$ might as well not be connected.
An odd-cycle is possibly retrieved after removing such repetitions, by inspecting the vertices in the sequence. Note that the remaining sequence may not be a closed walk, and the shortest path computation for this $u$ yields no cycle in $\hat{G}$ for the current solution \textbf{x}.

The weight of any such odd-cycle $U \subset C$ in the conflict graph, disregarding any chords it might have, is 
$w(U) = \sum_{\{i,j\} \in U} w(i,j) = \sum_{\{i,j\} \in U} \frac{(1 - x_i - x_j)}{2} = \frac{|U|}{2} - \frac{1}{2} \sum_{\{i,j\} \in U} (x_i + x_u) = \frac{|U|}{2} - \sum_{i \in V(U)} x_i$, where $V(U) \subseteq E$ denotes the set of nodes induced by $U$.
That is, $\sum_{i \in V(U)} x_i =  \frac{|U|}{2} - w(U)$, implying that \textbf{x} violates the corresponding odd-cycle inequality $\sum_{i \in V(U)} x_i  \leq \frac{|U| - 1}{2} = \frac{|U|}{2} - \frac{1}{2}$ if and only if $w(U) < \frac{1}{2}$.



\subsubsection*{Implementation details}

We highlight that, in the special case of an integral solution \textbf{x} $\in \mathbb{B}^{m}$, there can be no violated odd-cycle inequality, since the edge inequalities (\ref{mp:stab1}) guarantee a stable set in $\hat{G}$.
Hence, we try to find a violated SEC. In this case, we need reduced asymptotic complexity than in the fractional case:
\textbf{x} is feasible in $P_{sec}$ if and only if the corresponding edges induce exactly one connected component in $G$, which we check in $O(m)$ time with DFS. If there are different components, inspecting them yields the subtour and the inequality to add, and we terminate the procedure without considering the above algorithms to separate a fractional \textbf{x}.

\cite{RebennackTutorial2012} also indicate some implementation tweaks for the separation of odd-cycle inequalities. Among those, we include two simple and effective adjustments in our approach.
First, the auxiliary graph $H$ can be greatly reduced by removing nodes $u^+$ and $u^-$ whenever $x_u$ is integer, since no odd-cycle including $u$ could yield a violated constraint, as we explain next.
Suppose $U$ is an odd-cycle in $\hat{G}(E,C)$, and that it includes $j \in E$, with $x_j = 0$. Now, 
$\sum_{i \in U} x_i =  \sum_{i \in U, i \neq j} x_i \leq \frac{|U \backslash \left\lbrace j \right\rbrace|}{2} = \frac{|U| - 1}{2}$, where the inequality holds because every two consecutive nodes can contribute at most $1$ to the sum in the left hand side, provided edge inequalities (\ref{mp:stab1}) are satisfied.
Similarly, $x_j = 1$ implies two neighbors in $U$ with null value, and the above argument applies.
The second refinement regards the case of \textbf{x} such that $x_u + x_v = 1$ for a given $(u,v) \in C$. The definition of the weight function $w$ provides that both edges $(u^+, v^-)$ and $(u^-, v^+)$ in $H$ would have null cost. It is more interesting, though, to add a small weight $\epsilon$ instead, to avoid unnecessary vertices in the shortest path. We use $\epsilon = 10^{-6}$, as the authors suggest.

Finally,  preliminary experiments indicated that different strategies for reinforcing the relaxed polyhedron when separating the current solution have a major impact on computational performance. Standard strategies include returning as soon as a first cut is found, looking for the most violated inequality, or including all violated cuts.
We could verify the best overall results with an alternative strategy looking for \emph{some} of the best cuts: we include not only the most violated inequality, but also others 
which are close enough to being orthogonal to it. Pilot studies suggested accepting hyperplanes with an inner product of $0.1$ or less.
This enhanced strategy seems to balance the strength and diversity of included cuts, allowing to solve the LP relaxation in time similar to that of including all violated cuts, while limiting the model size.



\section{Computational results}
\label{sec:results}

The goals of the computational evaluation we present are twofold: to assess the impact of the preprocessing algorithm and the strengthening inequalities from $P_{stab}$, and to indicate how stronger are the bounds we provide than those previously available in the literature.
To the best of our knowledge, these correspond to the Lagrangean relaxation scheme of \cite{Zhang2011}, where a maximum edge clique partitioning subproblem is solved. As for the primal bounds reported by such authors, we consider the best result achieved by one of the heuristics they propose.

The algorithm we describe is implemented in \textit{C++}, using the callback mechanism in the Concert API of CPLEX 12.5. We turn off all preprocessing, heuristics and cut generation options -- only user cuts are separated. We consider a numerical precision of $10^{-5}$, even when looking for violated constraints.
Experiments were carried on a machine with an Intel Core i7 980 (3.33GHz) CPU, with 24GB of RAM. We use the benchmark instances proposed by \cite{Zhang2011}; as these are integer-valued, we set the absolute MIP gap tolerance parameter of CPLEX to $0.9999$.
We refer to an instance defined on a graph $(V,E)$ and conflict set $C$ by the identifier $|V|-|E|-|C|$.
We set an overall (wall-clock) time limit of $5000$ seconds.
Note that \cite{Zhang2011} used a different but comparable experimental setup. Their experiments were conducted on a Dell PC with 3.40 GHz Intel Pentium processor and 2.0 GB memory running Windows XP operating system and a Dell workstation with a 2.0 GHz Intel Xeon processor and 512 MB of memory running the Linux operating system. They used CPLEX 9.1 on the workstation to solve the integer programs and also used an overall (wall-clock) time limit of $5000$ seconds.

The benchmark includes \emph{type 1} and \emph{type 2} instances.
The first set includes harder problems, and several instances have neither optimality nor feasibility certificates available.
The latter set is much easier in practice, possibly because its instances are made feasible through a heuristic.
In fact, the preprocessing algorithm had major impact on this set, as its denser conflict graphs are more amenable to the probing techniques we apply.
Table \ref{tab:preprocessing} presents the effectiveness of the algorithm with type 1 (upper section of the table) and type 2 problems (lower section).
The fourth column indicates the total number of edges fixed, while the fifth reports the number of new conflict pairs included in the last phase of the algorithm. The next column indicates the resulting instance dimensions, or the certificate provided (when that is the case), followed by the total execution time.

\begin{table}[t]
\centering
\tiny
\caption{Instance reduction using the preprocessing algorithm.}
\begin{tabular}{ c c c | c c c c}
    \hline
	$|V|$	&	$|E|$	&	$|C|$	&	\# \textbf{Edges Fixed}	&	\# \textbf{Conflicts Included}	&	\textbf{Resulting Instance}	&	\textbf{Time (s)}	\\
    \hline
	50	&	200	&	199	&	0	&	0	&	$ |V|=50, |E|=200, |C|=199 $	&	0.05	\\
	50	&	200	&	398	&	0	&	0	&	$ |V|=50, |E|=200, |C|=398 $	&	0.05	\\
	50	&	200	&	597	&	0	&	0	&	$ |V|=50, |E|=200, |C|=597 $	&	0.06	\\
	50	&	200	&	995	&	0	&	11	&	$ |V|=50, |E|=200, |C|=1006 $	&	0.13	\\
	100	&	300	&	448	&	0	&	23	&	$ |V|=100, |E|=300, |C|=471 $	&	0.4	\\
	100	&	300	&	897	&	1	&	135	&	$ |V|=100, |E|=299, |C|=1026 $	&	0.77	\\
	100	&	300	&	1344	&	1	&	188	&	$ |V|=100, |E|=299, |C|=1472 $	&	0.9	\\
	100	&	500	&	1247	&	0	&	0	&	$ |V|=100, |E|=500, |C|=1247 $	&	0.79	\\
	100	&	500	&	2495	&	0	&	0	&	$ |V|=100, |E|=500, |C|=2495 $	&	0.86	\\
	100	&	500	&	3741	&	0	&	2	&	$ |V|=100, |E|=500, |C|=3743 $	&	1.81	\\
	100	&	500	&	6237	&	0	&	31	&	$ |V|=100, |E|=500, |C|=6268 $	&	1.99	\\
	100	&	500	&	12474	&	8	&	2747	&	$ |V|=100, |E|=492, |C|=12720 $	&	13.8	\\
	200	&	600	&	1797	&	0	&	126	&	$ |V|=200, |E|=600, |C|=1923 $	&	3.96	\\
	200	&	600	&	3594	&	0	&	504	&	$ |V|=200, |E|=600, |C|=4098 $	&	7.34	\\
	200	&	600	&	5391	&	--	&	--	&	\textbf{Infeasible}	&	9.05	\\
	200	&	800	&	3196	&	0	&	6	&	$ |V|=200, |E|=800, |C|=3202 $	&	7.75	\\
	200	&	800	&	6392	&	0	&	27	&	$ |V|=200, |E|=800, |C|=6419 $	&	8.46	\\
	200	&	800	&	9588	&	0	&	175	&	$ |V|=200, |E|=800, |C|=9763 $	&	8.93	\\
	200	&	800	&	15980	&	1	&	1220	&	$ |V|=200, |E|=799, |C|=16558 $	&	55.44	\\
	300	&	800	&	3196	&	--	&	--	&	\textbf{Infeasible}	&	41.42	\\
	300	&	1000	&	4995	&	0	&	201	&	$ |V|=300, |E|=1000, |C|=5196 $	&	46.32	\\
	300	&	1000	&	9990	&	1	&	661	&	$ |V|=300, |E|=999, |C|=10477 $	&	42.04	\\
	300	&	1000	&	14985	&	--	&	--	&	\textbf{Infeasible}	&	60.23	\\
	\hline													
	50	&	200	&	3903	&	159	&	1	&	$|V|=33, |E|=41, |C|=12 $	&	0.05	\\
	50	&	200	&	4877	&	167	&	3	&	$|V|=27, |E|=33, |C|=10 $	&	0.03	\\
	50	&	200	&	5864	&	175	&	1	&	$|V|=21, |E|=25, |C|=7 $	&	0.09	\\
	100	&	300	&	8609	&	287	&	0	&	$|V|=12, |E|=13, |C|=0 $	&	0.05	\\
	100	&	300	&	10686	&	291	&	0	&	$|V|=9, |E|=9, |C|=0 $	&	0.03	\\
	100	&	300	&	12761	&	291	&	0	&	$|V|=9, |E|=9, |C|=0 $	&	0.06	\\
	100	&	500	&	24740	&	464	&	35891	&	$|V|=32, |E|=36, |C|=2 $	&	2.25	\\
	100	&	500	&	30886	&	469	&	0	&	$|V|=28, |E|=31, |C|=0 $	&	0.24	\\
	100	&	500	&	36827	&	465	&	0	&	$|V|=33, |E|=35, |C|=1 $	&	0.21	\\
	200	&	400	&	13660	&	368	&	0	&	$|V|=30, |E|=32, |C|=1 $	&	0.01	\\
	200	&	400	&	17089	&	382	&	0	&	$|V|=17, |E|=18, |C|=1 $	&	0.01	\\
	200	&	400	&	20469	&	392	&	0	&	$|V|=8, |E|=8, |C|=0 $	&	0.01	\\
	200	&	600	&	34504	&	567	&	0	&	$|V|=32, |E|=33, |C|=0 $	&	0.59	\\
	200	&	600	&	42860	&	584	&	0	&	$|V|=16, |E|=16, |C|=0 $	&	0.19	\\
	200	&	600	&	50984	&	588	&	0	&	$|V|=12, |E|=12, |C|=0 $	&	0.09	\\
	200	&	800	&	62625	&	785	&	0	&	$|V|=14, |E|=15, |C|=0 $	&	0.29	\\
	200	&	800	&	78387	&	755	&	0	&	$|V|=42, |E|=45, |C|=0 $	&	0.24	\\
	200	&	800	&	93978	&	786	&	0	&	$|V|=14, |E|=14, |C|=0 $	&	0.69	\\
	300	&	600	&	31000	&	--	&	--	&	\textbf{Optimal}	&	0.45	\\
	300	&	600	&	38216	&	555	&	0	&	$|V|=44, |E|=45, |C|=1 $	&	0.02	\\
	300	&	600	&	45310	&	575	&	0	&	$|V|=23, |E|=25, |C|=0 $	&	0.02	\\
	300	&	800	&	59600	&	795	&	0	&	$|V|=5, |E|=5, |C|=0 $	&	0.03	\\
	300	&	800	&	74500	&	775	&	0	&	$|V|=25, |E|=25, |C|=0 $	&	0.03	\\
	300	&	800	&	89300	&	780	&	0	&	$|V|=20, |E|=20, |C|=0 $	&	0.04	\\
	300	&	1000	&	96590	&	984	&	0	&	$|V|=16, |E|=16, |C|=0 $	&	2.08	\\
	300	&	1000	&	120500	&	--	&	--	&	\textbf{Optimal}	&	9.12	\\
	300	&	1000	&	144090	&	--	&	--	&	\textbf{Optimal}	&	17.18	\\
    \hline
\label{tab:preprocessing}
\end{tabular}

\end{table}

All type 2 instances become trivial problems, with an empty conflict set in most cases, resulting in a standard MST problem.
The branch and cut algorithm systematically solves them to optimality in the root LP relaxation node,  executing in negligible wall clock time.
We therefore present no further results for these instances.
Note, however, that considerably large duality gaps are left by the algorithms discussed by \cite{Zhang2011}.

On the other hand, no similar effect was verified for type 1 instances. In most cases, no edge could be fixed at all, even though the conflict graph could be extended by the pair probing technique of the last phase in the algorithm.
Moreover, actually solving the resulting models indicated a minor impact on solution bounds, in comparison to solving the original instance without preprocessing.

Finally, note that six instances are solved during the preprocessing phase. Three type 2 instances are reduced to a problem defined on a tree without conflicting edges, in which case the solution is unique.
Three type 1 instances were proved to be infeasible.
Interestingly, while the branch and cut algorithm could also prove two of the problems to be infeasible within the time limit, the certificate for the largest one (\begin{small}$300-1000-14985$\end{small}) was only provided by the preprocessing algorithm, which is always executed in the remaining evaluations.


We consider next the impact of using the constraints obtained from the polytope of stable sets in the conflict graph. 
Table \ref{tab:sec:improvement} presents the percentual improvement on dual bounds over the plain formulation without these inequalities: i.e. columns \emph{OCI} correspond to the impact of intersecting the spanning tree polytope with the cycle-constrained relaxation $P_{cstab}$ instead of the simplest relaxation $P_{rstab}$; analogously, columns \emph{Cliques} use $P_{qstab}$ instead of $P_{rstab}$, while \emph{OCI+Cliques} compare $P_{qstab} \cap P_{cstab}$ with $P_{rstab}$. We report on percentual strengthening on both the initial LP relaxation and on the final bound provided by the branch and cut algoritm, using all type 1 instances which are not proved to be infeasible.

\begin{table}[t]
\centering
\scriptsize
\caption{Impact of odd-cycle and clique inequalities on dual bounds.}
\label{tab:sec:improvement}
\begin{tabular}{c | c c c | c c c}
    \hline
	\multirow{2}{*}{\textbf{Instance}}  &  \multicolumn{3}{ c | }{\textbf{LP Relaxation Bound}}  &  \multicolumn{3}{ c }{\textbf{MIP Lower Bound}} \\
	&	OCI	&	Cliques	&	OCI+Cliques	&	OCI &	Cliques	&	OCI+Cliques	\\
    \hline
	$50-200-199$	&	0.0	&	0.0	&	0.0	&	0.0	&	0.0	&	0.0	\\
	$50-200-398$	&	1.7	&	1.4	&	1.4	&	0.1	&	0.0	&	0.0	\\
	$50-200-597$	&	2.3	&	0.7	&	2.0	&	0.1	&	0.0	&	0.1	\\
	$50-200-995$	&	22.9	&	7.1	&	23.3	&	0.0	&	0.0	&	0.0	\\
	$100-300-448$	&	0.0	&	0.0	&	0.0	&	0.0	&	0.0	&	0.0	\\
	$100-300-897$	&	6.2	&	2.4	&	6.4	&	0.0	&	0.0	&	0.0	\\
	$100-300-1344$	&	18.3	&	6.5	&	18.5	&	-1.5	&	2.8	&	-0.9	\\
	$100-500-1247$	&	0.0	&	0.0	&	0.1	&	0.0	&	0.0	&	0.0	\\
	$100-500-2495$	&	6.6	&	1.6	&	6.6	&	0.0	&	0.0	&	-0.7	\\
	$100-500-3741$	&	17.0	&	7.6	&	17.0	&	2.6	&	4.6	&	2.7	\\
	$100-500-6237$	&	30.4	&	23.0	&	30.8	&	13.5	&	15.9	&	14.6	\\
	$100-500-12474$	&	44.1	&	62.2	&	62.8	&	6.3	&	19.8	&	14.1	\\
	$200-600-1797$	&	3.7	&	0.7	&	3.7	&	-0.5	&	-0.1	&	-1.2	\\
	$200-600-3594$	&	26.2	&	10.5	&	26.2	&	14.2	&	7.2	&	16.5	\\
	$200-800-3196$	&	3.2	&	0.8	&	3.4	&	-0.9	&	0.5	&	-0.4	\\
	$200-800-6392$	&	24.1	&	9.1	&	24.0	&	18.6	&	8.5	&	18.8	\\
	$200-800-9588$	&	38.5	&	31.2	&	39.2	&	26.5	&	24.2	&	26.4	\\
	$200-800-15980$	&	48.1	&	52.6	&	53.7	&	31.8	&	46.0	&	37.8	\\
	$300-1000-4995$	&	15.8	&	3.3	&	15.8	&	12.5	&	2.8	&	12.0	\\
	$300-1000-9990$	&	33.8	&	21.8	&	33.8	&	26.9	&	17.0	&	25.4	\\
	\hline
	Average	&	17.2	&	12.1	&	18.4	&	7.5	&	7.5	&	8.3	\\
    \hline
\end{tabular}

\end{table}

In general, OCI contributes more to tightening the LP dual bound. Still, the complete formulation using both classes is never worse (disregarding a factor of $0.1\%$), as expected.
The point supporting the methodology of using the strongest formulation is clearer for the final branch and cut bounds.
On average, OCI and clique inequalities provide equivalent contribution, though one of them is remarkably more effective in each particular instance. The intuition on using both of them is therefore to capture each scenario in the proposed algorithm.
In this sense, the complete formulation (with \emph{OCI+Cliques}) performs better, on average, than isolated counterparts. It is worth noting that the improvement is greater for larger problem instances.


Finally, we indicate how stronger are the bounds provided by the present approach, comparing them with the best results described by \cite{Zhang2011}.
Table \ref{tab:zhang:open} presents the results of the branch and cut algorithm with formulation (\ref{mp:mstcc}), considering all type 1 instances.
The third column indicates LP relaxation bounds, while the fourth presents \emph{$[$primal, dual$]$} bounds provided by branch and cut. The last column depicts our percentual improvement on the best dual bound previously available.

Two new feasibility certificates are provided, yielding the first primal bounds on instances \begin{small}$200-600-1797$\end{small} and \begin{small}$200-800-3196$\end{small}, as well as the
the new optimality certificate for instance \begin{small}$100-300-897$\end{small}
(in bold on Table \ref{tab:zhang:open}).
We also note that the improvement on previous results varies interestingly among instances, ranging from $14\%$ to $89\%$, motivating further work and different approaches for the MSTCC problem.

A key result in this work regards the consistent improvement of previous known dual bounds already by the initial LP relaxation bound of the proposed formulation. On average, the bounds on the third column are $25\%$ stronger than those achieved by the Lagrangean lower bounding scheme of \cite{Zhang2011} after hours of computation; in fact, they report execution times of up to $28421.5$ seconds.

\begin{table}[t]
\centering
\scriptsize
\caption{Comparing results with \cite{Zhang2011} on type 1 instances.}
\label{tab:zhang:open}
\begin{tabular}{c c | c c c}
	\hline
	\multirow{2}{*}{\textbf{Instance}}	&	\textbf{Bounds of}	&	\multicolumn{3}{ c }{\textbf{Branch and cut with (\ref{mp:mstcc})}} \\
		&	\textbf{\cite{Zhang2011}}	&	LP Bound	&	MIP Bounds	&	Improvement (\%)	\\
	\hline
	$50-200-199$	&	[702.793 , 708]	&	701	&	708	&	0.7	\\
	$50-200-398$	&	[757.816 , 785]	&	758	&	770	&	1.6	\\
	$50-200-597$	&	[807.745 , 1044]	&	852.9	&	917	&	13.5	\\
	$50-200-995$	&	[877.495 , 1424]	&	1179.2	&	1324	&	50.9	\\
	$100-300-448$	&	[3991.18 , 4102]	&	3991.3	&	4041	&	1.2	\\
	$100-300-897$	&	[4624.24 , --]	&	5196.8	&	\textbf{5658}	&	22.4	\\
	$100-300-1344$	&	[4681.27 , --]	&	6059.3	&	[6621.2 , --]	&	41.4	\\
	$100-500-1247$	&	[4165.68 , 4293]	&	4247.6	&	4275	&	2.6	\\
	$100-500-2495$	&	[4805.40 , 6603]	&	5557	&	[5951.4 , 6006]	&	23.8	\\
	$100-500-3741$	&	[4871.27 , 8787]	&	6259.4	&	[6510.8 , 9440]	&	33.7	\\
	$100-500-6237$	&	[4968.99 , --]	&	7189.2	&	[7568.7 , --]	&	52.3	\\
	$100-500-12474$	&	[5194.67 , --]	&	9011.5	&	[9816.9 , --]	&	89.0	\\
	$200-600-1797$  	&	[11425.8 , --]    	&	12906.2	&	[13072.9 , \textbf{14707.0}]	&	14.4	\\
	$200-600-3594$	&	[12487 , --]	&	16791.5	&	[17532.7 , --]	&	40.4	\\
	$200-600-5391$	&	[12873.2 , --]	&	--	&	\textbf{infeasible}	&	--	\\
	$200-800-3196$  	&	[17992.6 , --]    	&	20303.2	&	[20744.2 , \textbf{21852.0}]	&	15.3	\\
	$200-800-6392$	&	[19705.7 , --]	&	25929.1	&	[26361.3 , --]	&	33.8	\\
	$200-800-9588$	&	[20684.8 , --]	&	29230	&	[29443.6 , --]	&	42.3	\\
	$200-800-15980$	&	[20226.9 , --]	&	32271.9	&	[33345.1 , --]	&	64.9	\\
	$300-800-3196$	&	[30190.1 , --]	&	--	&	\textbf{infeasible}	&	--	\\
	$300-1000-4995$	&	[40732.7 , --]	&	51066.3	&	[51451.3 , --]	&	26.3	\\
	$300-1000-9990$	&	[42902.5 , --]	&	59884.6	&	[60907.8 , --]	&	42.0	\\
	$300-1000-14985$	&	[44639.1 , --]	&	--	&	\textbf{infeasible}	&	--	\\
	\hline									
\end{tabular}

\end{table}



\section{Final remarks}
\label{sec:final}

This work contributes with an exact solution approach to MSTCC.
We present a general preprocessing algorithm based on implications from feasibility conditions, which might be integrated to different methodologies for the problem.
We introduce an IP formulation building on classic polyhedral descriptions to represent the feasibility of a solution with respect to both input graphs $G(V,E)$ and $\hat{G}(E,C)$. Computational results with the preprocessing and a branch and cut algorithm indicate a consistent improvement on the best results previously available in the literature, and provide new feasibility and optimality certificates for benchmark instances of the problem.

Further research could elaborate on a specific analysis of the MSTCC polytope, and possibly indicate the form of valid inequalities, whether also valid for the stable set relaxation or not.  It would be interesting to know under which conditions are odd-hole and maximal clique inequalities also facet-defining for the new polytope. While such issues are interesting in their own right, the algorithmic approach may require enhanced formulations and to leverage more established techniques from the stable set literature, e.g. the balanced branching rule of \cite{balasYuClique1986}, or the edge projection technique to separate rank inequalities, a large class which generalize both odd-cycle and clique inequalities \citep{edgeProjection2001,RebennackTutorial2012}.

We also seek to gain insight into the hardness and tractability of related problems under disjunctive constraints, such as the ones described by \cite{damPTM2011}. Note that our approach might be appropriate for these problems as well.


\begin{acknowledgements}
The authors wish to thank the anonymous reviewers for carefully reading the ma\-nus\-cri\-pt and for the suggestions that helped improving our presentation.
Phillippe Samer is supported by a grant from CAPES (Coordenadoria de A\-per\-fei\-\c{c}o\-a\-men\-to de Pessoal de N\'{i}vel Superior, Brazil).
Sebasti\'{a}n Urrutia is partially supported by CNPq (Conselho Nacional de Desenvolvimento Cient\'{i}fico e Tecnol\'{o}gico, Brazil) grant 303442/2010-7. \end{acknowledgements}



\end{document}